\documentclass[aps,prl,superscriptaddress,amsmath,amssymb,a4paper,twocolumn,longbibliography]{revtex4-1}

\usepackage{graphicx}
\usepackage{dcolumn}
\usepackage{bm}
\usepackage{hyperref}
\usepackage{color}

\usepackage[shortlabels]{enumitem}

\usepackage{natbib}
\usepackage{comment}

\begin{document}


\title{Revealing polymerization kinetics with colloidal dipatch particles}

\author{Simon Stuij} \affiliation{Institute of Physics, University of Amsterdam, Science Park 904, 1098 XH Amsterdam, the Netherlands}
\author{Joep Rouwhorst} \affiliation{Institute of Physics, University of Amsterdam, Science Park 904, 1098 XH Amsterdam, the Netherlands}
\author{Hannah Jonas} \affiliation{van 't Hoff Institute for Molecular Sciences, University of Amsterdam, Science Park 904, 1098 XH Amsterdam, the Netherlands}
\author{Nicola Ruffino} \affiliation{Institute of Physics, University of Amsterdam, Science Park 904, 1098 XH Amsterdam, the Netherlands}
\author{Zhe Gong} \affiliation{New York University}
\author{Stefanno Sacanna} \affiliation{New York University}
\author{Peter Bolhuis} \affiliation{van 't Hoff Institute for Molecular Sciences, University of Amsterdam, Science Park 904, 1098 XH Amsterdam, the Netherlands}
\author{Peter Schall} \affiliation{Institute of Physics, University of Amsterdam, Science Park 904, 1098 XH Amsterdam, the Netherlands}



\begin{abstract}
Limited-valency colloidal particles can self-assemble into polymeric structures analogous to molecules. While their structural equilibrium properties have attracted wide attention, insight into their dynamics has proven challenging.
Here, we investigate the polymerization dynamics of semiflexible polymers in two dimensions (2D) by direct observation of assembling divalent particles, bonded by critical Casimir forces. The reversible critical Casimir force creates living polymerization conditions with tunable chain dissociation, association and bending rigidity. We find that unlike dilute polymers that show exponential size distributions in excellent agreement with Flory theory, concentrated samples exhibit arrest of rotational and translational diffusion due to a continuous isotropic-to-nematic transition in 2D, slowing down the growth kinetics. These effects are circumvented by addition of higher-valency particles, cross-linking the polymers into networks. Our results connecting polymer flexibility, polymer interactions and the peculiar isotropic-nematic transition in 2D offer insight into polymerization processes of synthetic two-dimensional polymers, and biopolymers at membranes and interfaces.
\end{abstract}

\maketitle

Continuous advances in colloidal synthesis endow colloids with a design space potentially as diverse as available to molecular structures \cite{Gong2017, Rogers2016, Kraft2012,Wolters2015,Evers2016, Sacanna2010, McMullen2018, BenZion2017, Shah2014, Yan2016}. These particles can provide model systems to elucidate structural and dynamical behavior of their molecular counterparts. Colloidal polymers specifically promise to bring the richness of molecular-, supramolecular-, and biopolymers to the colloidal scale, with the benefit that particles can be conveniently observed and interaction potentials can be externally controlled.

While a broad theoretical understanding of polymers has developed starting from the foundational work of Flory \cite{Flory1953},
many open questions remain, including a predictive framework for non-equilibrium polymerisation \cite{Tantakitti2016}, and polymerization in confinement or low dimensionality \cite{Jordens2013,Czogalla2015,Lackinger2015,Colson2013}.
Biologically relevant semiflexible polymers such as DNA, microtubules or amyloids often polymerize in confinement or at membranes, and at high density, undergo alignment into a nematic phase, affecting their effective stiffness, polymerization kinetics, and biological function.
While in three dimensions (3D), the nematic ordering is a well-understood first-order transition, in two dimensions (2D), due to the prevalence of fluctuations in low-dimensional systems, the transition is typically of second order or continuous Kosterlitz-Thouless type, depending on the particle interactions and rigidity \cite{Jordens2013}. The alignment affects the diffusional properties of the filaments \cite{Hore2010,Czogalla2015}, which in turn can affect the polymerization process itself \cite{Luiken2015}, but the effect of filament rigidity and interactions on the nature of the 2D isotropic-nematic transition and ultimately on the 2D polymerization kinetics remain unclear.

\begin{figure*}[t!]
    \includegraphics[width=15cm]{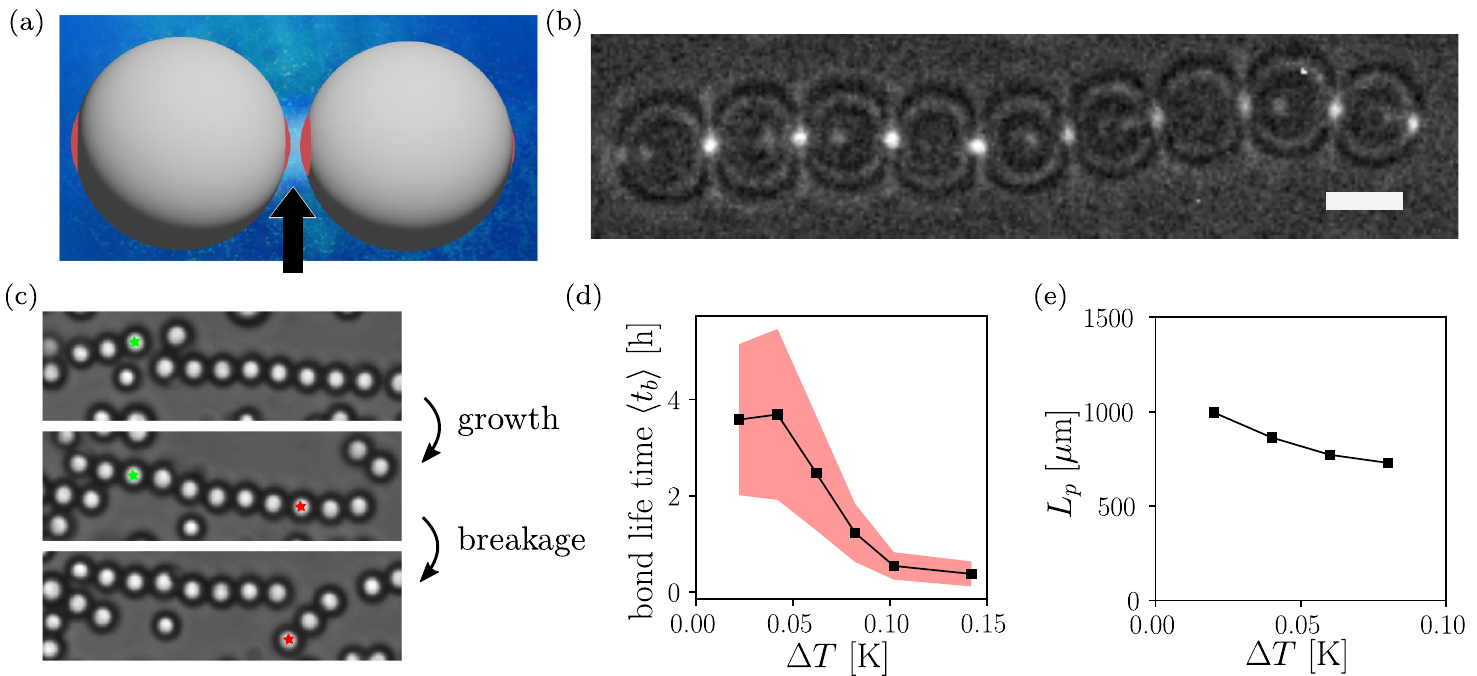}
    \caption{{\bf Polymerization of colloidal di-patch particles by critical Casimir forces} (a) Illustration of di-patch particles with lutidine-affine patches confining lutidine-rich fluctuations in the binary mixture (b) Epifluorescence microscopy image of a single colloidal polymer chain, highlighting the fluorescently dyed patches (bright dots) binding the particles. Scale bar is $3~\mathrm{\mu m}$. (c) Sequential bright-field images taken at $\Delta T = 0.06 \mathrm{K}$ showing attachment (green dots) and detachment events (red stars). (d) Average bond life time versus $\Delta T$, determined by monitoring of individual chains. Shaded area represents uncertainty. (e) Persistence length $L_p$ versus $\Delta T$ obtained from bending fluctuations~\cite{Stuij2020}. Values correspond to normalized persistence length $L_\mathrm{p}/d$ between 230 and 310.
    }
    \label{fig:figure1}
\end{figure*}

A powerful model system to obtain generic insight are self-assembling patchy particles that achieve limited valency with only a minimal adjustment, a surface with well-defined sticky patches of tunable size \cite{Zhang2004,Duguet2016,Bianchi2017}. Simulations have shown that dipatch particles polymerize in exact agreement with Flory theory \cite{Sciortino2007,Russo2010}, and in more dense cases exhibit an isotropic-to-nematic phase transition \cite{Lue2004,DeMichele2012,Nguyen2018}. Experimental studies are scarce \cite{Wang2012,McMullen2018,Naderi2020}, and often limited by irreversibility of the assembly, or the inability to grow larger structures. Endowing patchy particles with critical Casimir interactions \cite{Soyka2008,Nellen2009, Mohry2014, FarahmandBafi2020} opens new opportunities for reversible directed assembly: the confinement of fluctuations of a near-critical binary solvent between the surface-modified patches (Fig.~\ref{fig:figure1}(a)) causes an attractive patch-patch interaction that is in a universal manner set by temperature.


Here we use reversible bonding of di-patch particles by critical Casimir forces to investigate low-dimensional polymerization. We vary the bond strength and find that at sufficiently low concentration, chain association and dissociation lead to exponential size distributions consistent with Flory theory. At higher concentration, the growing semiflexible chains align with nearby chains, emerging into a continuous isotropic-to-nematic transition. The concomitant slow-down of rotational diffusion leads to kinetic arrest, resulting in lower growth rates and shorter chains. We show that the alignment is circumvented by including higher-valency (tetrapatch) particles, which cross-link the chains into a space-spanning network characterized by power-law size distributions. Our system provides a platform to test theories of polymer physics in a directly observable way. It paves the way towards complex colloidal structures assembled using finely tuned patchy particle interactions.

Dipatch and tetrapatch particles are synthesized from polystyrene (bulk material) and fluorescently labeled 3-(trimethoxysilyl)propyl methacrylate (TPM, surface patches) using a recently published colloidal fusion method~\cite{Gong2017}. The particles have diameters $d_\mathrm{2p}=3.2\mathrm{\mu m}$ and $d_\mathrm{4p}=3.7~\mathrm{\mu m}$ respectively, and patch sizes $\theta_\mathrm{2p}=21\mathrm{^{\circ}}$ and $\theta_\mathrm{4p}=16\mathrm{^{\circ}}$, with $\theta_\mathrm{2/4p}$ the arc angle of the spherical patch, as determined by atomic force microscopy. They are dispersed in a binary lutidine-water mixture with lutidine volume fraction $c_\mathrm{L}=0.25$, just below the critical concentration $c_\mathrm{c} = 0.30$~\cite{Stein1972}.
We add $0.375$mM $\mathrm{MgSO_4}$ salt to screen particle charges and enhance the lutidine adsorption of the patches. Suspensions are prepared at particle volume fractions $0.005$, $0.010$, and $0.020 \%$, corresponding to surface coverages of $\approx20$, $30$, and $60\%$ after particle sedimentation. To study equilibrium polymerization, the dilute sample is heated to $\Delta T = 0.02$K below $T_{c} = 33.86^{\circ}\mathrm{C}$, left to assemble for $10$ hours, and cooled in steps of $0.02$K until $\Delta T = 0.14$K to reduce the critical Casimir attraction, while equilibrating for $4$ hours at each temperature. To study the polymerization kinetics, the samples are heated to $\Delta T = 0.06$K and left to assemble for $15$ hours. We finally study cross-linked polymers by adding tetrapatch to the dipatch particles at number fractions of $0.1$ and $0.05$.


Upon approaching $T_c$, dipatch particles attach via their patches, polymerizing into chains as shown in Fig.~\ref{fig:figure1}(b). The critical Casimir bonding is fully reversible and tunable with temperature. Frequent attachment and breakage events are observed (Fig.~\ref{fig:figure1}(c)), with a bond lifetime $t_\mathrm{b}$ shortening with increasing $\Delta T$, as shown in Fig.~\ref{fig:figure1}(d). The finite bond lifetime creates living polymerization conditions, in which monomer and oligomer growth and detachment are present.
Concomitantly, the chain persistence length $L_p$, determined from bending fluctuations, decreases from $1000\mathrm{\mu m}$ to $750\mathrm{\mu m}$ (Fig.~\ref{fig:figure1}e), a factor $\sim 5$ shorter than that of microtubules, $L_p \approx 5000~\mathrm{\mu m}$ \cite{Gittes1993}.

\begin{figure}[t!]
    \includegraphics[width=\columnwidth]{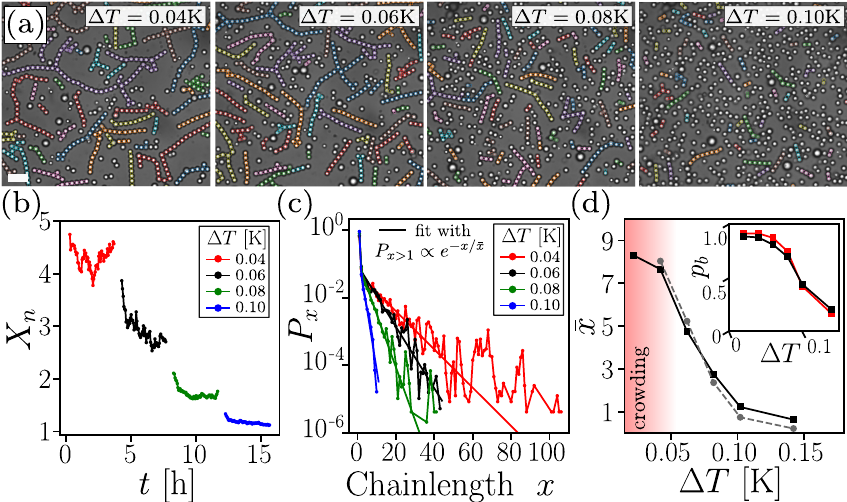}
    \caption{{\bf Equilibrium polymerisation} (a) Microscope images of dipatch particles polymerized at decreasing interaction strength (increasing $\Delta T$) and surface coverage $\phi=0.28$. Overlain colored circles mark individual chains. Scale bar is $25\mathrm{\mu m}$ (b) Average chain length ($X_n$) versus time, color indicates temperature steps of $0.02$K (c) Corresponding chain length distributions. Lines are exponential fits. (d) Fitted characteristic length $\bar{x}$ versus $\Delta T$ of experiments (black squares) and simulations (gray circles). Red shade indicates non-equilibrium region. Inset: corresponding bonding probability according to Flory theory (black), and extracted from observed bonding time $t_b$ using $1/t_b \propto (1-p_b)$  (red).}
    \label{fig:figure2}
\end{figure}

The decreasing bond strength gives rise to shorter polymers, as shown for the equilibrated samples in Fig.~\ref{fig:figure2}(a). To quantify this observation, we identify connected chains and determine their lengths $x$ and frequency of lengths $N_x$. The average chain length $X_n = \frac{\sum x N_x }{N}$, with $N$ the total number of chains, saturates at decreasing values for increasing $\Delta T$ (Fig.~\ref{fig:figure2}(b)), suggesting equilibration at the lower attractive strengths.
Length distributions $P_x=N_x/N$ are shown in Fig.~\ref{fig:figure2}(c); apart from an excess of monomers, the data follow exponential distributions with attractive-strength dependent exponent: as the attraction increases, the decay flattens, indicating longer chains. This becomes most obvious when we fit the distributions (excluding monomers) with $P_x \propto e^{-x/\bar{x}}$ (solid lines in Fig.~\ref{fig:figure2}(c)), and plot the resultant characteristic length $\bar{x}$ versus temperature in Fig.~\ref{fig:figure2}(d). In Flory theory, each binding site has an equally likely bonding probability $p_b$ \cite{Flory1953, Sciortino2007}, leading to the exponential chain length distribution $P_x = (1-p_\mathrm{b})p_\mathrm{b}^{x-1}$. Here, $p_\mathrm{b}$ is related to the characteristic chain length $\bar{x}$ via $p_\mathrm{b} = e^{-1/\bar{x}}$. Alternatively, we can determine $p_\mathrm{b}$ directly from the measured bonding time $t_\mathrm{b}$ (Fig. ~\ref{fig:figure1}(d)) using $t_\mathrm{B}^{-1} \propto (1-p_\mathrm{b})$. Indeed, we can scale both on top of each other as shown in Fig.~\ref{fig:figure2}(d) inset, confirming the agreement with Flory theory predictions.

These observations are confirmed in Monte Carlo simulations~\cite{Jonas2020} with an optimized potential based on critical Casimir scaling theory~\cite{Stuij2017} and an angular switching function mimicking the particle patches in experiments. The simulations reproduce the experimental chain length distributions very well: An excess of singlets is followed by an exponential distribution for longer chains, data shown in SI. A good agreement between simulations and experiment is observed in the fitted characteristic length $\bar{x}$ as a function of temperature, see Fig.~\ref{fig:figure2}(d). This agreement further confirms that the experimentally observed polymerisation is in equilibrium.

\begin{figure}[ht!]
    \includegraphics[width=\columnwidth]{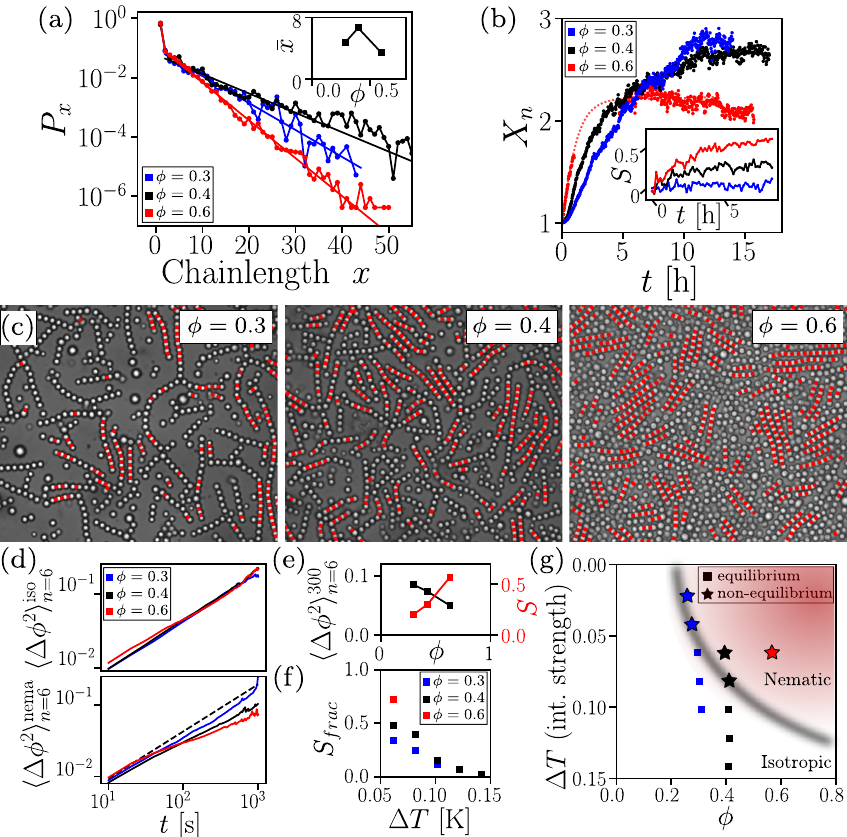}
    \caption{{\bf Colloidal polymerization kinetics.} (a) Chain length distributions for increasing surface coverage at $\Delta T = 0.06$ after $15$h of aggregation, inset shows fitted $\bar{x}$ (b) Average degree of polymerization, $X_n$ versus time. Inset: average local nematic order $S$ versus time (c) Microscope images at advanced stage of polymerization for increasing surface coverage. Red dots indicate local nematic order $S(i)>0.4$. (d) Mean-square rotation $\langle \Delta \phi^2 \rangle^{iso}_{x=6}$ of bonds in isotropic (top) and nematic environment (bottom), for chains of length $x=6$. Dashed line indicates power-law fit to data in top. (e) Mean-square rotation after $\Delta t=300$s (black) and average local nematic order $S$ (red) versus surface coverage.
    (f) Nematic ratio as a function of $\Delta T$ for all experiments in their final assembly state. The nematic ratio, $S_{frac}=N^b_{S>0.4}/N^b$ is the fraction of the total number of bonds, $N^b$, with $S>0.4$. (g) Phase diagram estimated from experimental data. The red shade indicates the nematic region. Data points indicate equilibrium (squares) and nonequilibrium states (stars).
    }
    \label{fig:figure3}
\end{figure}

At surface coverages of 0.4 and 0.6, the distributions deviate increasingly from the equilibrium ones, as shown by the non-monotonic behavior in Fig.~\ref{fig:figure3}a. Chains first become longer for higher monomer concentration, but then shorten. We identify a kinetic origin behind this trend, see Fig.~\ref{fig:figure3}(b): While initially, the polymers grow faster for higher monomer concentration as expected, at a later stage, the more concentrated samples slow down and become overtaken by the less concentrated ones.

To investigate the dynamic slow-down in more detail, we look at the structure of the polymerizing samples. In dense samples, the effective steric interaction tends to align the chains, as clearly visible in the snapshots in Fig.~\ref{fig:figure3}(c). We quantify the local alignment of bond $i$ with unit vector $\mathbf{u}_i$ with respect to the neighboring bonds $j$ with unit vectors $\mathbf{u}_j$ using the local nematic order parameter in two dimensions~\cite{Frenkel1985,Cuetos2007,Karner2020} $ S_i = \frac{1}{N_b} \sum^{N_b}_{j=0} 2 ||\mathbf{u}_i \cdot \mathbf{u}_j ||^2-1$, where $N_\mathrm{b}$ is the number of neighbors of bond $i$, defined as those particles having their bond centers closer than $2d_\mathrm{2p}$. The red dots highlighting aligned bonds ($S(i)>0.4$) show increasing nematic order at increasing density. This nematic order also grows in time as the samples polymerize, as shown in Fig.~\ref{fig:figure3}(b) inset.

The propagating alignment slows down the rotational dynamics, as can be seen by comparing the rotational diffusion in isotropic and nematic environments in Fig.~\ref{fig:figure3}(d). In isotropic environment, rotation follows a subdiffusive power law, with power $\sim 0.7$ independent of particle concentration, whereas in nematic regions, rotational diffusion slows down with increasing particle concentration. This is because nematic clusters have to rotate as a whole, which is much slower than the rotation of a single chain. The concomitant slowdown of rotational diffusion and increase in nematic order with particle density are clearly shown in Fig.~\ref{fig:figure3}(e).

The alignment effects suggest that the system wants to undergo an isotropic-to-nematic phase transition \cite{Chen1993}. Indeed, the fraction of aligned bonds ($S_{frac}$) increases with attraction and particle density, see Fig.~\ref{fig:figure3}(f).
This is in agreement with simulations of dipatch particles that show the existence of a nematic phase for high enough density and interaction strength \cite{Hentschke1989,Lue2004,DeMichele2012,Nguyen2018}. However, while the isotropic-to-nematic transition is of first order in 3D, the continuous increase of the order parameter in Fig.~\ref{fig:figure3}(f) and the gradual transition from near-nematic to isotropic areas in Fig.~\ref{fig:figure3}(c) suggest that in our 2D case, the transition is of rather continuous type, as observed in simulations of 2D hard rods \cite{CosentinoLagomarsino2003}; yet, the exact classification of the transition needs a rigorous analysis of an equilibrium state. This has dramatic consequences for the polymerization: while theory and simulation studies in 3D predict a sharp {\it increase} of the polymer chain length when the system undergoes the transition to a nematic phase~\cite{Hentschke1989,Lue2004,DeMichele2012}, we observe the opposite: a {\it decrease} of chain length when chains align. We associate this qualitatively different behavior with the more continuous nature of the isotropic-to-nematic transition in 2D, causing frustration of intertwined regions and kinetic arrest. 
It appears therefore that precisely when the system undergoes the isotropic-to-nematic transition, it becomes kinetically arrested and ends up trapped in a non-equilibrium state.

These findings are summarized in the schematic diagram in Fig.~\ref{fig:figure3}(g). At low attraction and monomer concentration, the equilibrium state is isotropic; with increasing attraction strength and concentration, the system undergoes a transition to the nematic phase, marked by the red shade. The symbols indicate equilibrium polymer growth (squares) and  out-of-equilibrium configurations (stars), as defined by the deviation from the expected equilibrium size distribution. These results highlight the intriguing relation between the polymerization kinetics and the isotropic-to-nematic transition.

Reconstituted microtubule systems undergo nematic ordering in 3D \cite{Hitt1990}, 2D~\cite{CosentinoLagomarsino2003}, and when activated~\cite{Sanchez2012}. Furthermore, amiloyd fibrils show nonequilibrium isotropic-nematic coexistence when confined at liquid interfaces~\cite{Jordens2013}. While in the 3D and activated cases, no signs of crowding were observed, in 2D the system arranged in aligned domains similar to our case. Furthermore, simulations of amyloid fibril formation have shown that aligned peptide clusters diffuse more slowly, similar to our aligned dipatch polymer clusters \cite{Luiken2015}. Our controlled colloidal model allows direct particle-scale insight into the interplay of polymerization, emergence of nematic order, and kinetic arrest.


\begin{figure}[ht!]
    \includegraphics[width=\columnwidth]{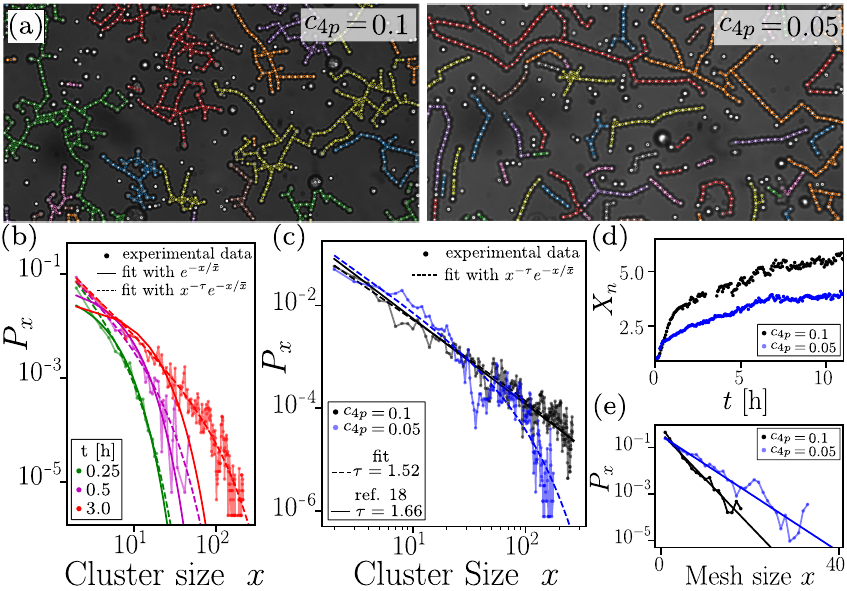}
    \caption{
    {\bf Crosslinked colloidal polymer networks.} (a) Microscope images of colloidal polymer networks obtained with a small fraction $c_{4p}=0.1$ (left) and $c_{4p}=0.05$ (right) of tetrapatch particles. Color indicates individual clusters. (b) Cluster size distributions at different times, with power-law fits ($\tau=-1.52$) with exponential cutoff $\bar{x}=4$, $11$, $68$ (dotted) (c) Final cluster size distribution after $10$h of assembly. Curves indicate power-law fit with exponent $\tau=-1.52$ and exponential cutoff $\bar{x}=50$ ($480$) for blue (black) datapoints. (d) Average cluster size $X_n$ versus time. (e) Distribution of strand lengths in between particles that are bonded with three or more particles.}
    \label{fig:figure4}
\end{figure}

The alignment and kinetic arrest are alleviated by addition of a small fraction of tetrapatch particles that offer additional binding sites, leading to cross-linked polymer networks, see Fig.~\ref{fig:figure4}(a). As the crosslinks form, the initial exponential cluster-size distribution develops a pronounced tail and switches to a power law, see Fig.~\ref{fig:figure4}(b) and (c). This is clearly reflected in the exponential fits (solid lines), which break down after $3$ hours of growth.
This power-law behavior indicates percolation of the network, and is markedly different from the exponential behavior of the purely linear polymers, as it breaks with the predictions of Flory theory \cite{Flory1953}. We find a power-law exponent of $\tau=1.52\pm 0.15$, consistent within fitting accuracy with 2D simulations that find $\tau=1.66$~\cite{Russo2010} (solid line in Fig.~\ref{fig:figure4}(c)). As was noted by Russo et al.~\cite{Russo2010} these $\tau$ values deviate from 2D perculation predictions $\tau = 187/91$. This is surprising because in 3D percolation theory did apply \cite{Bianchi2007}.

The changing topology appears to also reduce the kinetic arrest, as shown in Fig. \ref{fig:figure4}(d). Besides some slow down, the cluster size continues to rise (approximately linearly). Apparently, crowding effects are less prominent when including enough tetrapatch particles, due to the additional binding sites and different geometry. Indeed, the nematic state disappears upon including tetrapatch particles, as their inclined bond angles make alignment unfavorable. Finally, decreasing the tetrapatch (i.e. crosslinker) particle fraction changes the network strand length as shown in Fig. \ref{fig:figure4}(e), while it does not affect $\tau$. We can thus tune the network mesh size by the amount of tetrapatch particles.

Reversible assembly of limited-valency particles by critical Casimir forces opens up new opportunities for investigation of semiflexible colloidal polymers, polymer networks and possibly many more (bio)molecule-inspired structures. We find that at sufficient concentration, semiflexible colloidal polymers align, causing a gradual isotropic-to-nematic transition, and leading to slow down of the rotational diffusion and polymerization. 
These results provide insights into crowded biopolymers and active nematics, where similar alignment effects are observed. Comparisons to other self-assembling nematics can be made such as living liquid crystals \cite{Bladon1993}. The direct particle-scale vizualisation gives insight into polymerization processes of recent (disordered) 2D polymers \cite{Colson2013,Lackinger2015} whose synthesis is still challenging. Furthermore, the effect of kinetic trapping in the final non-equilibrium assembled state also lays the groundwork for modeling kinetic pathways and their role in desired final non-equilibrium states, a key recent development in polymer chemistry \cite{Tantakitti2016,Canning2016}.
Finally, by including active colloidal particles, this system could form well-controlled and conveniently observable active nematics, which currently attract a lot of attention for their rich non-equilibrium behavior \cite{Sanchez2012,Doostmohammadi2018} .

\begin{acknowledgments}
We thank Michele Zanini for the AFM data and analysis.
\end{acknowledgments}

\bibliography{PaperBibliography}

\end{document}